# High precision measurements of half-lives for $^{69}$Ge, $^{73}$Se, $^{83}$Sr, $^{85m}$Sr, and $^{63}$Zn radionuclides relevant to the astrophysical *p*-process via photoactivation at the Madison Accelerator Laboratory

T. A. Hain[1], S. J. Pendleton[1], J. A. Silano[2], and A. Banu[1*]

[1]Department of Physics and Astronomy, James Madison University, Harrisonburg, Virginia 22807, USA

[2]Nuclear and Chemical Sciences Division, Lawrence Livermore National Laboratory, Livermore, California 94550, USA

[*]Corresponding author: banula@jmu.edu

**Abstract**

The ground state half-lives of $^{69}$Ge, $^{73}$Se, $^{83}$Sr, $^{63}$Zn, and the half-life of the 1/2$^-$ isomer in $^{85}$Sr have been measured with high precision using the photoactivation technique at an unconventional bremsstrahlung facility that features a repurposed medical electron linear accelerator. The γ-ray activity was counted over about 6 half-lives with a high-purity Germanium detector, enclosed into an ultra low-background lead shield. The measured half-lives are: $T_{1/2}$($^{69}$Ge) = 38.82 ± 0.07 (stat) ± 0.06 (sys) h; $T_{1/2}$($^{73}$Se) = 7.18 ± 0.02 (stat) ± 0.004 (sys) h; $T_{1/2}$($^{83}$Sr) = 31.87 ± 1.16 (stat) ± 0.42 (sys) h; $T_{1/2}$($^{85m}$Sr) = 68.24 ± 0.84 (stat) ± 0.11 (sys) min; $T_{1/2}$($^{63}$Zn) = 38.71 ± 0.25 (stat) ± 0.10 (sys) min. These high-precision half-life measurements will contribute to a more accurate determination of corresponding ground-state photoneutron reaction rates, which are part of a broader effort of constraining statistical nuclear models needed to calculate stellar nuclear reaction rates relevant for the astrophysical *p*-process nucleosynthesis.

## Introduction

Nuclear physics is fundamental to two major themes in astrophysics – nucleosynthesis of the chemical elements by nuclear reactions in stars and stellar explosions and energy production in such environments. Although there is a solid understanding of the production of elements up to iron by nuclear fusion reactions in stars, important details concerning the production of the elements beyond iron remain puzzling. Current knowledge is that the nucleosynthesis beyond Fe proceeds mainly *via* neutron capture reactions and subsequent β$^-$ decays in the so-called *s*-process and *r*-process [1]. However, some 35 *proton-rich stable* isotopes between $^{74}$Se and $^{196}$Hg cannot be synthesized by neutron-capture processes as they are located on the neutron-deficient side of the valley of β-stability thus they are shielded from *s*- or *r*-processes. These proton-rich nuclides are generally referred to as *p*-nuclei [2] and the mechanism responsible for their synthesis is termed the *p*-process. As a group the *p*-nuclei are the rarest of all stable nuclei, their solar and isotopic abundances being typically one or two orders of magnitude lower than the respective *s*- and *r*-process nuclei. The main process of production of *p*-nuclei is the destruction of pre-existing *s*- or *r*-nuclei by sequences of photodisintegrations, such as (γ,n), (γ,p) or (γ,α) reactions, and β$^+$ decays.

Modelling of the *p*-process nucleosynthesis requires an extended network of more than 20000 reactions linking more than 2000 nuclei in the A ≤ 210 mass region [3]. It is impossible to measure all these reaction cross sections in the laboratory and then extract their reaction rates. Hence, it becomes obvious that the vast majority of the relevant *p*-process reaction cross sections and reaction rates must be calculated, and usually it is done within the framework of Hauser-Feshbach (HF) statistical models [3]. Such models require input based on nuclear structure physics, optical model potentials, nuclear level densities, and γ-ray strength functions which in turn determine the reaction cross sections and thus their corresponding reaction rates.

The stellar photodisintegration reaction rates relevant for the *p*-process nucleosynthesis calculations are dominated by excited state contributions at the high temperature regime of the *p*- process. Thus, they cannot be directly studied experimentally, because in the laboratory one mainly has access to target nuclei in the ground state. Therefore, the goal of photodisintegration reaction cross section measurements is to provide instead nuclear input to constrain crucial parameter of the statistical nuclear reaction models, i.e. γ-ray strength function.

In line with this goal, we were motivated to investigate several photoneutron reactions on stable nuclei relevant to the *p*-process nucleosynthesis via photoactivation technique using bremsstrahlung photons from an electron linear accelerator by applying the so-called superposition method [4]. This experimental technique has been established and well tested allowing the determination of the ground-state photodisintegration reaction rates directly from bremsstrahlung-induced activation experiments. We report in this paper on high precision measurements of half-lives for $^{69}$Ge, $^{73}$Se, $^{83}$Sr, $^{85m}$Sr, and $^{63}$Zn radionuclides, produced very selectively via the $^{70}$Ge(γ,n), $^{74}$Se(γ,n), $^{83}$Sr(γ,n), $^{86}$Sr(γ,n), and $^{64}$Zn(γ,n) reactions, respectively. The progenitor nuclei $^{70}$Ge, $^{74}$Se, $^{84}$Sr, $^{86}$Sr and $^{64}$Zn are proton-rich stable nuclei that belong to the region A < 124, notoriously underproduced by the current stellar evolution models for the astrophysical *p*-process. As the decay parameters enter into the experimental determination of the ground-state photoneutron rates of interest, their precise values must be known.

Photoactivation studies using bremsstrahlung as real photon sources have been widely performed at conventional linear accelerators around the world [5], most notably at the superconducting Darmstadt linear electron accelerator S-DALINAC [6] and at the superconducting electron accelerator ELBE of the Forschungszentrum Rossendorf [7]. However, using an "off-the-shelf" medical linear acclerator, originally designed for clinical operation and repurposing it for nuclear physics experiments, is a novel idea. We have successfully explored it in this work in light of a pioneering photoactivation study [8] and of more recent photoneutron average cross section measurements [9-12], all carried at clinical electron linacs, demonstrating that the excellent stability and reproducibility of the beam of this type are good pre-requisites for photonuclear experiments.

**Experimental Details**

The half-lives of five proton-rich radioisotopes that belong to a mass region relevant to the astrophysical *p*-process – $^{69}$Ge, $^{73}$Se, $^{83}$Sr, $^{85m}$Sr, and $^{63}$Zn - were measured with high precision via photon-induced activation at the Madison Accelerator Laboratory (MAL), a unique bremsstrahlung facility on the campus of James Madison University, in Harrisonburg, Virginia, USA. Newly operational since 2017, the facility features a medical electron linear accelerator, an X-ray imaging system, and a suite of particle detection instrumentation. The linear accelerator, a magnetron-powered Siemens Mevatron, generates a 6-μs-long pulsed electron beam at roughly 200 Hz pulse repetition frequency, which is impinged on a tungsten target producing bremsstrahlung photons with endpoint energies from 6 to 15 MeV.

Highly enriched samples of stable nuclei of $^{70}$Ge, $^{74}$Se, $^{84}$Sr, and $^{64}$Zn were irradiated, one at a time, with bremsstrahlung photons at an endpoint energy of 15 MeV for 10 min on/5 min off intervals for nine cycles for a total beam exposure time of 90 min over a span of 130 min (66% duty cycle); this operation was required to maintain stable operation of the accelerator and avoid overheating of its components. The samples were positioned on the photon beam axis at 52 cm from the tungsten radiator target to ensure exposure to a maximum photon flux, and were activated by (γ,n) reactions. The respective neutron threshold energies of the $^{70}$Ge(γ,n), $^{74}$Se(γ,n), $^{83}$Sr(γ,n), $^{86}$Sr(γ,n), $^{64}$Zn(γ,n) reactions - 11.533 MeV, 12. 058 MeV, 8.859 MeV, 11.492 MeV, 11.863 MeV – are all smaller than the linac's 15 MeV-endpoint energy used in this work. Hence, it was possible to produce the radionuclides of interest, $^{69}$Ge, $^{73}$Se, $^{83}$Sr, $^{85m}$Sr, and $^{63}$Zn.

The $^{70}$Ge, $^{74}$Se, $^{84}$Sr, and $^{64}$Zn samples were produced by Trace Sciences International and assembled in glass vials, see Table 1 for details on their physical and chemical characteristics.

**Table 1** Physical and chemical characteristics of the samples of interest.

| sample | element weight (mg) | Enrichment (%) | Isotopic composition (%) |
|---|---|---|---|
| Germanium metal | 1,000.0 | $^{70}$Ge 95.85 ± 0.20 | $^{72}$Ge 4.09, $^{73}$Ge 0.04, $^{74}$Ge 0.02 |
| Selenium metal | 1,000.0 | $^{74}$Se 98.20 ± 0.10 | $^{76}$Se 1.80 ± 0.10, $^{77,78,80,82}$Se < 0.10 |
| Strontium carbonate (SrCO$_3$) | 100.0 | $^{84}$Sr 76.40 ± 1.10 | $^{86}$Sr 4.64, $^{87}$Sr 1.96, $^{88}$Sr 17.0 |
| Zinc metal | 1,000.0 | $^{64}$Zn 99.40 ± 0.10 | $^{66}$Zn 0.39, $^{67}$Zn 0.04, $^{68}$Zn 0.17, $^{70}$Zn < 0.01 |

After a cooling time, for most of the measurements of less than 3 minutes from sample irradiation, the corresponding produced activity was counted offline using a high-purity Germanium (HPGe) detector (model GC6020/ULB-GC), manufactured by Mirion Technologies, former Canberra, which at the 1.33 MeV $^{60}$Co line had a typical relative efficiency ≥ 60% and an energy resolution (FWHM) of 1.90 keV. The detector, which has a vertical slimline cryostat, was embedded by design into an ultra low-background lead shield (model 777A), also manufactured by Mirion Technologies/Canberra. The lead shield is a 15-cm-thick cylinder with a 2.5-cm inner layer being selected for reduced $^{210}$Pb content to about 20 Bq/kg and is constructed from materials carefully selected for low background. To show the efficiency of the lead shield, Fig. 1 illustrates the very low γ-ray background from the detected radioactivity and its composition, which is dominated by 511 keV photons. The background data in Fig. 1 was measured for 8 hours, yielding only 573 (± 7.30%) counts in the area of the 511 keV photopeak.

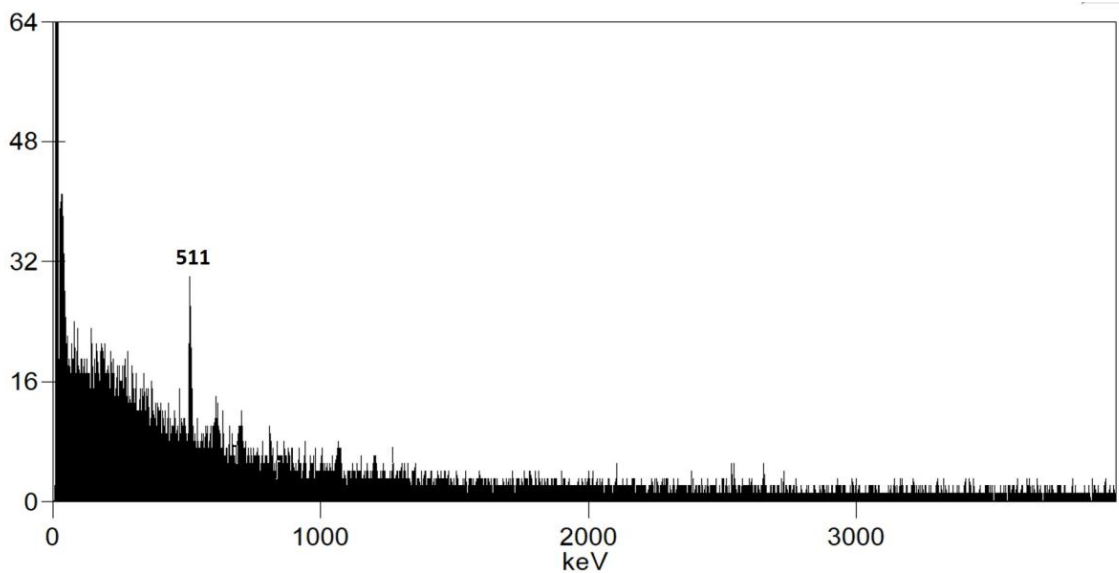

**Fig. 1** γ-ray energy calibrated spectrum of the background inside the lead shield chamber that hosted the HPGe detector used for the half-life measurements in this work (see text for details).

For signal processing and data acquisition, we used the Lynx digital multichannel analyzer (MCA) with Genie 2000 basic gamma spectroscopy software, both produced by Mirion Technologies/Canberra. The energy and efficiency calibrations of the collected γ-ray energy spectra were carried out using a calibrated multinuclide standard source [13], which contained the $^{241}$Am, $^{109}$Cd, $^{57}$Co, $^{139}$Ce, $^{203}$Hg, $^{113}$Sn, $^{85}$Sr, $^{137}$Cs, $^{88}$Y, and $^{60}$Co radionuclides. As the HPGe detector was vertically fitted inside the ultra low-background lead shield, it allowed placing easily and keeping the photoactivated samples or the calibration mixed source at a fixed location directly on the face of the detector endcap. Hence, we were able to control the reproducibility of the sample-detector geometry for all the offline sample decay counting, assuring that it could not become a source of systematic uncertainty in the sample half-life measurements. Trials were conducted with photoactivated sample spacing off the endcap and it was found that the activity of the samples was low enough that pulse pile up was not observed even with the sample resting directly on the endcap.

For each of the radionuclides of interest, the γ-ray emission spectra of the photoactivated samples were measured over a time period approximately equal to 6 half-lives. In order to obtain multiple independent measurements, and thus ensuring a more precise measurement of the half-life, each sample was irradiated and measured at least twice.

## Data Analysis and Results

In order to extract the half-lives of the radionuclides of interest, the corresponding γ-ray emission spectra were saved automatically at regular time intervals, generating 200-500 files throughout each measurement. A custom script was written in Visual Basic to interface with the Genie 2000 Gamma Acquisition & Analysis window to order the "save file" operation. The saved files contained information about the counts in each MCA channel of the γ-ray energy spectra, as well as the real and live time of the data acquisition.

When possible, more than one prominent γ-ray emission line was used in the data analysis for the same radionuclide. Below we list known energies [14] of the γ-ray lines used in this work to determine the half-life for each of the radionuclides of interest:

(i) **$^{69}$Ge**: 574.11 keV, 871.98 keV, 1106.77 keV
(ii) **$^{73}$Se:** 361.2 keV
(iii) **$^{83}$Sr:** 762.65 keV
(iv) **$^{85m}$Sr**: 151.19 keV
(v) **$^{63}$Zn:** 669.93 keV, 962.02 keV

As per the isotopic composition of the sample presented in Table 1, the strontium carbonate sample contained 4.64% $^{86}$Sr, which was most probably responsible for the production of the $^{85m}$Sr isomer via the respective (γ,n) reaction. We were not able to measure the half-life of the $^{85}$Sr radionuclide, as we could not disentangle the only prominent γ-ray transition at 514 keV (Iγ = 96%) [14], following the EC decay of the $^{85}$Sr into its daughter nucleus $^{85}$Rb, from the 511 keV photopeak.

To extract the half-lives using the net count areas of the photopeaks corresponding to each of the γ-ray lines listed above, the γ-ray spectra have been analyzed with the program package ROOT [15]. The net area of each photopeak of interest has been calculated by taking the difference between the integral of the measured γ-ray photopeak and the integral of the corresponding Compton background beneath each photopeak, over the region of interest (ROI) determined by side to side limits set in ROOT for each photopeak. The Compton background beneath each photopeak was evaluated by a background analytical function that was a fit to a region that was wider than the respective ROI of each photopeak but included the ROI within it. The background fit function was defined as a combination of a linear continuum and Gaussian peaks, since there were cases with nearby γ-ray peaks present in the vicinity of the photopeak of interest. An illustrative example of the Compton background subtraction procedure, implemented in our ROOT data analysis, is shown in Fig. 2, and corresponds to the γ-ray at 574.1 keV, which was one of the γ-ray lines used for the determination of the half-life of $^{69}$Ge.

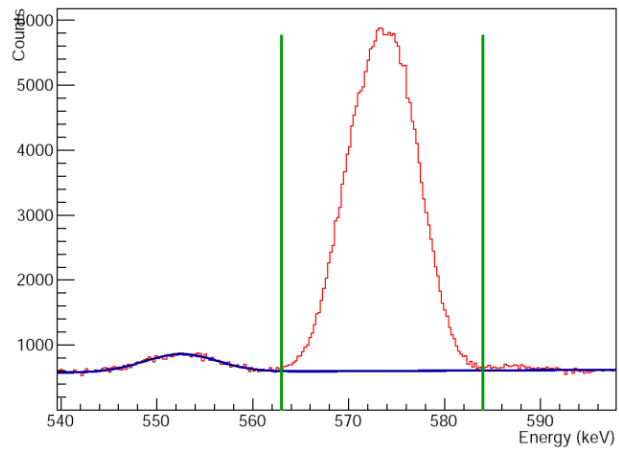

**Fig. 2** The γ-ray spectrum (red curve) corresponds to the measurement of the $^{69}$Ge decay, showing the γ-ray photopeak at 574.11 keV with the vertical green bars representing the software limits, selected in the ROOT data analysis as the ROI of the γ-ray photopeak of interest, and the blue line representing the background fit function (see text for details).

The net photopeak areas have been live-time corrected with a live time recorded to be 99% in all activity measurements from the first stored γ-ray spectrum to the last one for each of the nuclides of interest. In order to fit multiple time evolving spectra, the full series of spectra corresponding to the decay data have been summed and fit to determine the mean and spread values for the Gaussian photopeaks. These parameters were subsequently fixed for fitting individual spectra of interest, while the Gaussian amplitudes and the linear background were free fit parameters. The net photopeak areas for each time-binned spectrum were then determined by integrating the measured spectrum counts over the ROI and subtracting the background fit function over that same ROI, using the procedure described in the previous paragraph.

In order to check the consistency of the data analysis procedure applied for the substraction of the Compton background beneath each photopeak of interest, which constitutes the main source of systematic uncertanties in our half-life determination, ten different regions to fit the Compton background were chosen along with variations of the photopeak ROIs. Hence, the *systematic uncertanties* were evaluated as the standard deviation of the mean for each set of 10 different half-life measurements corresponding to each of the γ-ray photopeaks of interest.

The half-life results for $^{69}$Ge, $^{73}$Se, $^{83}$Sr, $^{85m}$Sr, $^{63}$Zn have been obtained from the linear fitting of the measured decay curves, plotted as the natural logarithm of the measured count rate as a function of time. Figs. 3-7 show a few samples of such plots along with the respective residual (%) plots to illustrate the goodness of the fit to the data.

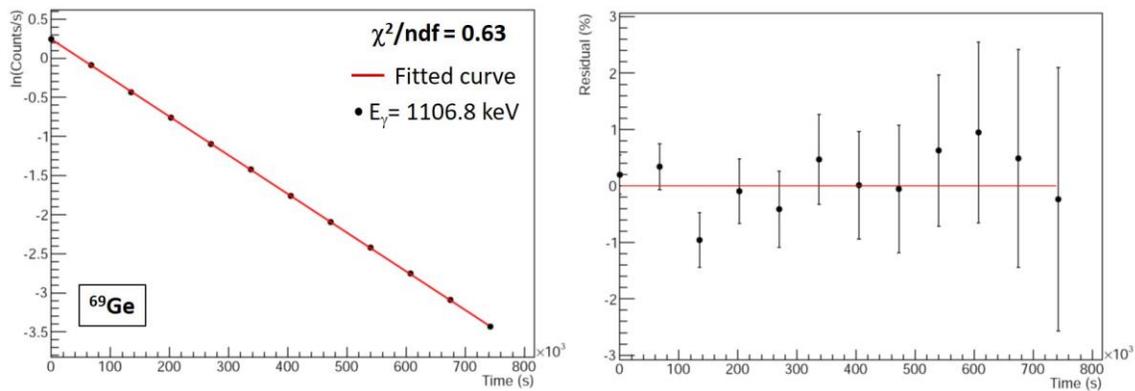

**Fig. 3** Decay curve of $^{69}$Ge at Eγ = 1106.8 keV (left) and its corresponding % residual between the linear fit and decay data points (right).

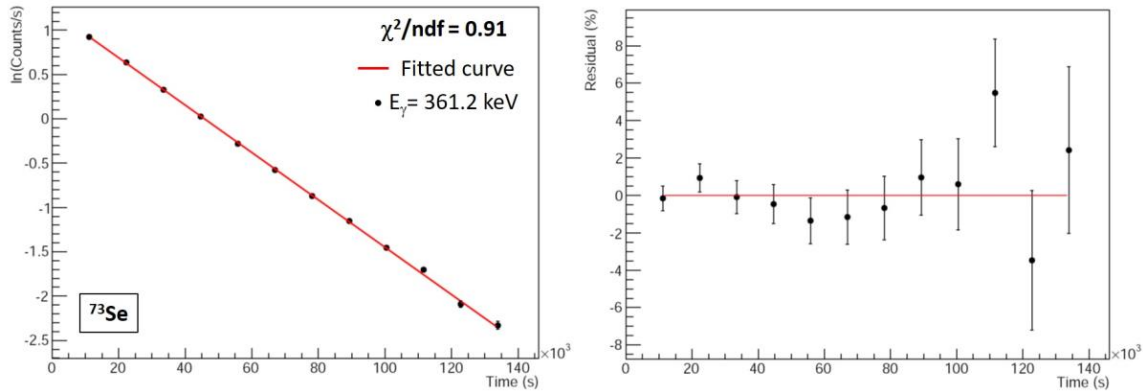

**Fig. 4** Decay curve of $^{73}$Se at Eγ = 361.2 keV (left) and its corresponding % residual between the linear fit and decay data points (right).

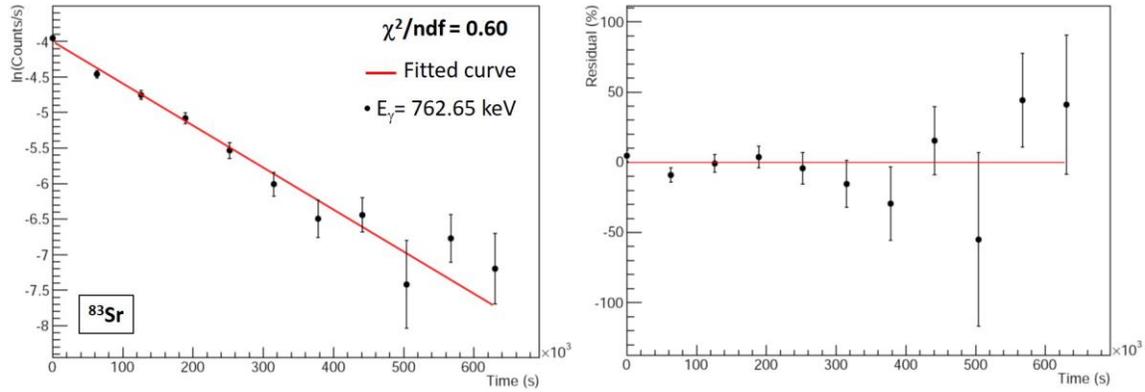

**Fig. 5** Decay curve of $^{83}$Sr at E$\gamma$ = 762.65 keV (left) and its corresponding % residual between the linear fit and decay data points (right).

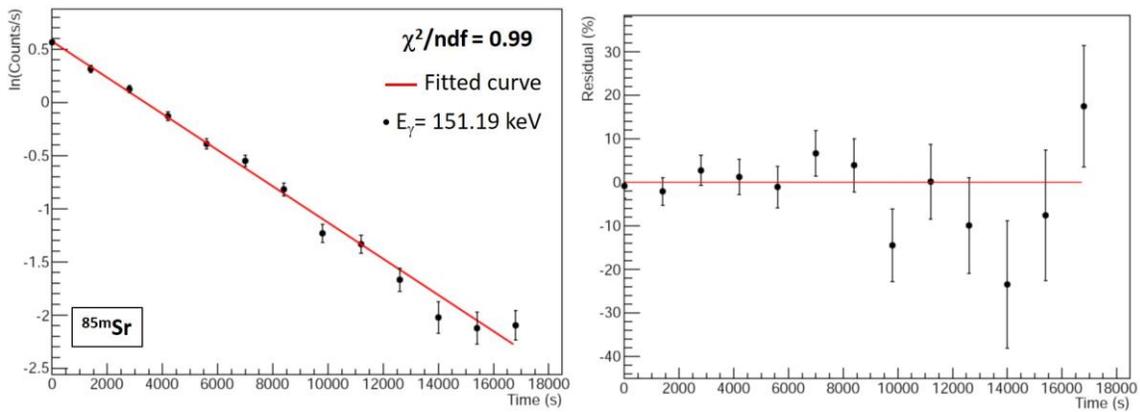

**Fig. 6** Decay curve of $^{85m}$Sr at E$\gamma$ = 151.19 keV (left) and its corresponding % residual between the linear fit and decay data points (right).

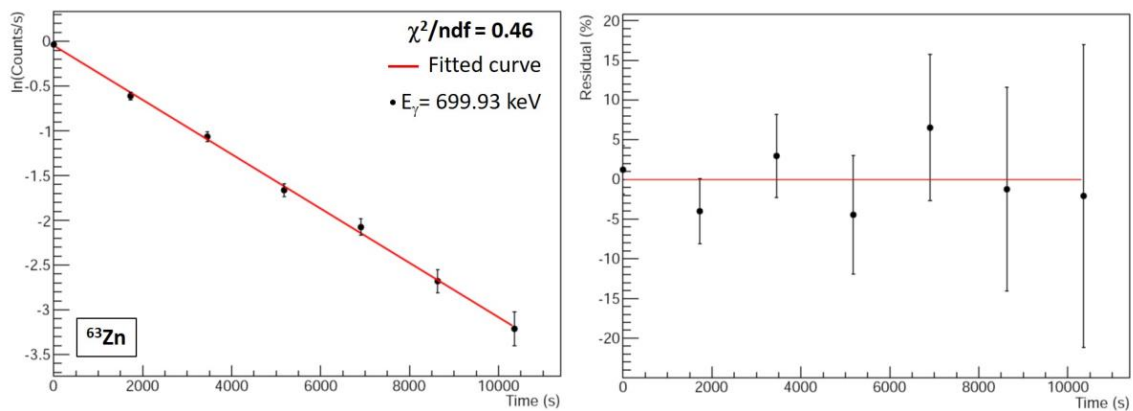

**Fig. 7** Decay curve of $^{63}$Zn at E$\gamma$ = 699.93 keV (left) and its corresponding % residual between the linear fit and decay data points (right).

It was mentioned at the end of the previous section that in order to extract more precise half-life values from the photoactivation data, we irradiated and measured each sample at least two times. Moreover, it was possible to measure simultaneously the respective half-life from more than one γ-ray lines. In light of this experimental approach, we treated our half-life results as values obtained from multiple independent measurements.

We present in Table 2 the individual half-life values as determined for each of the five radionuclides of interest from independent decay curve measurements from the observed γ-ray lines. These half-life values and their *statistical uncertainties* were calculated as the averages of each of the aforementioned sets of 10 different half-life measurements corresponding to ten different regions to fit the Compton background for each of the γ-ray photopeaks of interest. The *half-life weighted averages*, along with their statistical and systematic uncertainties, are shown in Table 2 in italic font style. The corresponding statistical uncertainties of the weighted avearges were calculated applying the standard formula:

$$\Delta \overline{T_{1/2}} = \sqrt{\frac{1}{\Sigma_i w_i}} \qquad (1)$$

where the weighting factors are the reciprocal of the squared statistical uncertainty of each result.

**Table 2** Half-life measurement results and other nuclear spectroscopic information from the present work along with the adopted values from the last ENSDF compilations, refs. [17-21]. See text for details on the determination of the listed statistical and systematic uncertainties. Half-life weighted averages are shown in italic font style.

| Isotope | Decay mode | $E_\gamma$ (keV) | $I_\gamma$ (%) | $T_{1/2}$ (present work) | $T_{1/2}$ (literature) |
|---|---|---|---|---|---|
| $^{69}$Ge | 100% $\beta^+$/EC | 574.11 | 13.3 | 38.68 ± 0.13 (stat) ± 0.02 (sys) | **39.05 ± 0.10 h** [17] |
|  |  | 871.98 | 11.9 | 38.76 ± 0.23 (stat) ± 0.03 (sys) |  |
|  |  | 1106.77 | 36 | 38.90 ± 0.09 (stat) ± 0.01 (sys) |  |
|  |  |  |  | *38.82 ± 0.07 (stat) ± 0.04 (sys)* |  |
| $^{73}$Se | 100% $\beta^+$/EC | 361.2 | 97.0 | 7.27 ± 0.07 (stat) ± 0.010 (sys) | **7.15 ± 0.09 h** [18] |
|  |  |  |  | 7.13 ± 0.04 (stat) ± 0.003 (sys) |  |
|  |  |  |  | 7.15 ± 0.03 (stat) ± 0.001 (sys) |  |
|  |  |  |  | 7.21 ± 0.03 (stat) ± 0.001 (sys) |  |
|  |  |  |  | *7.18 ± 0.02 (stat) ± 0.004 (sys)* |  |
| $^{83}$Sr | 100% $\beta^+$/EC | 762.65 | 26.7 | 30.94 ± 1.40 (stat) ± 0.29 (sys) | **32.41 ± 0.03 h** [19] |
|  |  |  |  | 33.88 ± 2.06 (stat) ± 0.54 (sys) |  |
|  |  |  |  | *31.87 ± 1.16 (stat) ± 0.42 (sys)* |  |
| $^{85m}$Sr | 13.4% EC | 151.19 | 12.8 | 68.69 ± 1.60 (stat) ± 0.11 (sys) | **67.63 ± 0.04 min** [19] |
|  |  |  |  | 67.83 ± 1.48 (stat) ± 0.11 (sys) |  |
|  |  |  |  | 68.26 ± 1.32 (stat) ± 0.12 (sys) |  |
|  |  |  |  | *68.24 ± 0.84 (stat) ± 0.11 (sys)* |  |
| $^{63}$Zn | 100% $\beta^+$/EC | 669.93 | 8.2 | 38.71 ± 0.49 (stat) ± 0.06 (sys) | **38.47 ± 0.05 min** [20] |
|  |  | 962.02 | 6.5 | 38.72 ± 0.29 (stat) ± 0.08 (sys) |  |
|  |  |  |  | *38.71 ± 0.80 (stat) ± 0.10 (sys)* |  |

In the following, we summarize our results from bremsstrahlung-induced activation.

- **⁶⁹Ge:** The measured half-life of **38.82 ± 0.07 (stat) ± 0.04 (sys) h** was obtained by combining the decay measurements for three of the strongest γ-ray transitions – 574.11 keV, 871.98 keV, and 1106.77 keV – in the daughter nucleus ⁶⁹Ga. The *0.04 h* systematic uncertainty of the weighted half-life result was determined by adding quadratically the values, shown in Table 2, of the respective systematic uncertainties of the three independent half-life results, obtained from the decay of the respective aforementioned γ-ray lines. Our result is slightly lower than the the adopted value of 39.05 ± 0.10 h [17] from [23].

- **⁷³Se:** Our result of **7.18 ± 0.02 (stat) ± 0.00 (sys) h** was obtained by combining the decay measurements for the strongest γ-ray transition – 361.2 keV – in the daughter nucleus ⁷³As, whose decay was measured four times. In this case, the systematic uncertainty of the weighted half-life result was determined as the average of the values, shown in Table 2, of the respective systematic uncertainties of the four independent half-life results, obtained from the decay of only one γ-ray line at 361.2 keV. Although the excited state in ⁷³As corresponding to the de-excitation via the 361.2 keV γ-ray line is not populated directly by the β⁺/EC- decay of the 3/2⁻ low-lying isomer in ⁷³Se to ⁷³As, the ⁷³ᵐSe metastable state ($T_{1/2}$ = 39.8 min [13]) also decays with a branching ratio of 72.6% [13] via internal transition (IT) to the ground state of ⁷³Se. To avoid the interference with the determination of the half-life of ⁷³Se, we disregarded in the data analysis the decay data files that corresponded to about five half-lives of this isomeric decay mode. The present measurements agree very well with the adopted value of 7.15 ± 0.09 h [18], which was compiled as a weighted average of 7.18 ± 0.09 h [24] and 7.08 ± 0.15 s [25], but we were able to significantly improve on precision of previous measurements. That was possible because we have used a very clean ⁷³Se source produced via the ⁷⁴Se(γ,n) reaction with an enriched sample of ⁷⁴Se. Moreover, we have also used a state-of-the-art ultra-low background HPGe detector for the respective γ-ray decay counting. In previous measurements the production of ⁷³Se was hindered because of the very low natural abundance (0.86%) of the ⁷⁴Se residuals from (α,xn) reactions, which produced mixed sources with more complicated γ-ray spectra and time-dependent background measured with small Ge(Li) detectors, overall less performant in their efficiency and energy resolution.

- **⁸⁵ᵐSr and ⁸³Sr:** The present result of **68.24 ± 0.84 (stat) ± 0.11 (sys) min** was obtained by combining the decay measurements for the γ-ray transition of 151.19 keV in the daughter nucleus $^{85}$Rb, which was measured three times. Also in this case, the systematic uncertainty of the weighted half-life result was determined as the average of the values, shown in Table 2, of the respective systematic uncertainties of the three independent half-life results, obtained from the decay of only one γ-ray line at 151.19 keV. This result is in agreement with the adopted value of 67.63 ± 0.04 min [18] compiled as a weighted average of 67.55 ± 0.07 min [26], 67.66 ± 0.07 min [27,28], 67.92 ± 0.25 min [28], and 67.3 ± 0.3 min [29].

We were also able to measure the decay of the $^{83}$Sr radionuclide produced via the $^{84}$Sr(γ,n) reaction by combining two decay measurements for the most prominent γ-ray in the daughter nucleus $^{83}$Rb, at 762.65 keV. In this case, due to a very poor peak-to-background ratio attributed to low counting statistics, we measured a half-life of **31.87 ± 1.16 (stat) ± 0.42 (sys) h** with a poor uncertainty, but in agreement with the literature adopted value of 32.41 ± 0.03 h [19] from [26].

- **$^{63}$Zn:** Our result of **38.71 ± 0.25 (stat) ± 0.10 (sys) min** was obtained by combining the decay measurements for the two strongest γ-ray transitions – 669.93 keV and 962.02 keV – in the daughter nucleus $^{63}$Cu, each measured two times. The *0.10 h* systematic uncertainty of the weighted half-life result was determined by adding quadratically the values, shown in Table 2, of the respective systematic uncertainties of the two independent half-life results, obtained from the decay of the respective aforementioned γ-ray lines. The present result is in good agreement with the adopted value of 38.47 ± 0.05 min [20] from [30].

## Conclusions

We have measured with high precision the half-lives of $^{69}$Ge, $^{73}$Se, $^{84,85m}$Sr and $^{63}$Zn. While our results are in good agreement with the respective literature adopted values, we were able in the case of $^{73}$Se to improve significantly on the experimental accuracy.

These results enable a more accurate determination of related ground-state photoneutron reaction rates for $^{70}$Ge(γ,n), $^{74}$Se(γ,n), $^{83}$Sr(γ,n), $^{86}$Sr(γ,n), $^{64}$Zn(γ,n), which can be useful in constraining statistical nuclear models needed to calculate stellar nuclear reaction rates relevant for understanding the astrophysical *p*-process nucleosynthesis. Moreover, our results demonstrate the viability of a converted medical linac as a bremsstrahlung source for photoactivation experiments and building on these results we will expand on this use with further work.

## Acknowledgements

We are indebted to Professor H. J. Karwowski for his thorough reading and editorial comments on manuscript drafts. Additionally, we would like to thank J. E. Mayer for assisting in the initial data reduction and Dr. P. Mohr for reading and insightful commenting on the early manuscript draft. Last but not least, we highly appreciated the thorough reviews by the anonymous reviewers. A. B. and T. A. H. acknowledge support by the National Science Foundation through Grant No. Phys-1913258. J. A. S. acknowledges that this work was performed under the auspices of the US Department of Energy by Lawrence Livermore National Laboratory under the Contract No. DE-AC52-07NA27344.